\documentclass[hidelinks]{article}
\usepackage{arxiv}
\usepackage{cite}
\usepackage{amsmath,amssymb,amsfonts}
\usepackage{algorithmic}
\usepackage{graphicx}
\usepackage{textcomp}
\usepackage{enumerate}
\usepackage{xcolor}
\def\BibTeX{{\rm B\kern-.05em{\sc i\kern-.025em b}\kern-.08em
    T\kern-.1667em\lower.7ex\hbox{E}\kern-.125emX}}

\title{Generation of Optimized Solidity Code for Machine Learning Models using LLMs}

\author{%
  Nikumbh Sarthak Sham\\
  Indian Institute of Technology Kharagpur, India\\
  \texttt{sarthaknikumbh@gmail.com} 
  \And
Sandip Chakraborty\\
  Indian Institute of Technology Kharagpur, India\\
  \texttt{sandipchkraborty@gmail.com} 
  \And
  Shamik~Sural \\
  Indian Institute of Technology Kharagpur, India\\
  \texttt{shamik@cse.ac.in} 
}

\date{}

\renewcommand{\headeright}{}
\renewcommand{\undertitle}{}
\renewcommand{\shorttitle}{Generation of Optimized Solidity Code for Machine Learning Models using LLMs}

\begin{document}
\maketitle

\begin{abstract}
While a plethora of machine learning (ML) models are currently available, along with their implementation on disparate platforms, there is hardly any verifiable ML code which can be executed on public blockchains. We propose a novel approach named LMST that enables conversion of the inferencing path of an ML model as well as its weights trained off-chain into Solidity code using Large Language Models (LLMs). Extensive prompt engineering is done to achieve gas cost optimization beyond mere correctness of the produced code, while taking into consideration the capabilities and limitations of the Ethereum Virtual Machine. We have also developed a proof of concept decentralized application using the code so generated for verifying the accuracy claims of the underlying ML model. An extensive set of experiments demonstrate the feasibility of deploying ML models on blockchains through automated code translation using LLMs.
\end{abstract}

\section{Introduction}
In recent years, two apparently divergent fields, namely Blockchain and Machine Learning (ML), have emerged as transformative technologies that are not only reshaping industries across the globe but also touching many facets of human life. 
% First introduced as the foundation for cryptocurrencies like Bitcoin \cite{nakamoto2008bitcoin}, blockchain has evolved far beyond its original purpose. It provides a secure, decentralized, and tamper-proof ledger system that enables trusted data sharing without the need for intermediaries. This ability to store and verify data immutably has opened up new possibilities across various sectors
% , including finance, healthcare, supply chain management, and many others 
% \cite{8487348}. Research and applications of machine learning, on the other hand, have progressed dramatically over the last two decades. ML is extensively being used for developing practical software for computer vision, speech and natural language processing, robot control, etc. \cite{doi:10.1126/science.aaa8415}. 
% ML has revolutionized how data can be effectively utilized and has enabled systems to learn from large amounts of data and improve autonomously over time. 
Convergence of blockchain and machine learning presents an interesting opportunity to harness the strengths of both. However, such an integration faces several major hurdles due to some of their inherent constraints. The Ethereum Virtual Machine (EVM) is computationally constrained whereas machine learning models are notoriously computation intensive, often requiring execution on GPU servers. Owing to this, there are a number of challenges in attempting to deploy ML models on a blockchain as identified below. 
\begin{itemize}
    \item \textbf{Gas Limitation:}
    It is one of the major constraints of EVM. Each transaction on the Ethereum network requires gas. Gas limit restricts the amount of computation that can be performed in a single transaction. Such a limitation is particularly prohibitive even for moderately sized machine learning models.
    \item \textbf{Storage Constraints:}
    EVM has storage constraints with maximum allowed stack size being 1024 and the maximum size of the stack data member set at 256 bits. Storing large datasets on a blockchain is both expensive as well as inefficient. In contrast, machine learning models are usually trained on very large datasets.
    \item \textbf{Lack of Floating-Point Support:}
EVM does not support floating point operations. On the other hand, almost every ML model requires floating point computations and sometimes even double precision representations.
\end{itemize}

Despite the challenges identified above, there are several advantages of deploying a machine-learning model on the blockchain. Extensive proliferation of machine learning has led to a proportionate rise in concerns regarding the verifiability of their computational outputs. Privacy is also an important consideration. Blockchain provides a decentralized platform for accessing deployed models, which enhances trust among users. Deployment of ML models on a blockchain ensures that all transactions related to the model are immutably recorded, and provides transparency in the decision-making process of the model. 

ML has several applications in blockchain which include verifying the reliability of data \cite{10634433}, federated learning \cite{self_IEEETP}, and finding vulnerabilities in smart contracts \cite{10634473}\cite{10634424}. In Decentralized Finance, ML is utilized for credit risk assessment \cite{10634435}, fraud detection \cite{10634350} and finding arbitrage opportunities in cryptocurrencies \cite{10634339}. These use cases can be more beneficial if the ML models are deployed on blockchain.

To address the concerns identified above and for exploring the opportunities, we propose a methodology named as LMST (\underline{L}LM based \underline{M}L to \underline{S}olidity code \underline{T}ranslator) that automates the process of translating machine learning code written in PyTorch to Solidity code for smart contracts. Instead of writing a custom code translator, which is inherently not scalable, we show how code translation can be done using Large Language Models (LLMs). We further optimize the smart contracts in terms of the gas consumed by leveraging the use of targeted prompts to the LLM. 
% A thorough gas cost analysis is performed in order to assess the feasibility of deploying these models. 
A prototype decentralized application (dApp) has been developed for multi-class classification using the Solidity code generated by LMST and deployed on the Ethereum Sepolia test network. Users can use the dApp to upload any image and get it recognized, thus being able to verify the performance of the ML model.

\section{Related Work}

In this section, we discuss some of the prior work that attempt to bridge the gap between machine learning and blockchain. One such effort is DInEMMo \cite{8634320}, which is a framework for integrating AI with blockchain to enhance the development and incentivization of machine learning models. It allows users to create models or enhance existing models in enterprise settings and defines a pricing strategy. Incentive calculation to reward the contributors fairly is based on their involvement in model enhancement. 
% The framework allows us to continually update the models. 
DanKu \cite{kurtulmus2018trustlessmachinelearningcontracts} is another protocol that establishes a marketplace for exchanging machine learning models in an automated and anonymous manner for participants. 
% The marketplace enables skilled individuals to monetize their machine learning expertise while organizations can solicit solutions worldwide. 
This approach aims to incentivize the creation of better machine learning models and make AI more accessible to companies and agents. However, DanKu does not allow continual updating and collaborative training of machine learning models.

Harris and Waggoner \cite{8946257} initially proposed a framework that outlines a decentralized collaborative approach to machine learning using blockchain. It specifically targets the issues of data centralization and model accessibility. 
% The framework provides free public access and 
% inference by hosting the models on smart contracts. This open access aims to democratize AI by reducing reliance on proprietary datasets and centralized models. The system incorporates both financial and gamified incentives to encourage participation and simultaneously prevent manipulation of the models. In this framework, contributors can earn rewards for submitting high-quality data and face penalties for poor submissions. The framework aims to create a large and publicly available dataset by crowdsourced contributions. 
Further extending this, Harris \cite{10.1007/978-3-030-59638-5_10} conducted an in-depth analysis of the framework with major focus on the self-assessment incentive mechanism. Three machine learning models were included in the study, namely, Perceptron \cite{rosenblatt1958perceptron}, Naïve Bayes \cite{webb2010naive} and Nearest Centroid Classifier \cite{tibshirani2002diagnosis}. These models were tested across three distinct datasets, which include fake news detection \cite{fake_news_kaggle_2020}, user activity prediction dataset \cite{ni2019modeling}, and IMDB movie review sentiment analysis dataset \cite{maas2011learning}.

In contrast to the above, Kadadha \textit{et al.} \cite{KADADHA2022170} present a novel approach to behavior prediction for task allocation within a blockchain-based crowd sourcing framework. This work proposed an on-chain ML model that predicts worker behavior based on task context. The model is trained off-chain to minimize costs and then deployed on-chain as a smart contract, which enables transparent predictions of worker behavior. Badruddoja \textit{et al.} \cite{9922480} address the challenges of implementing machine learning algorithms within blockchain smart contracts. 
% Their work focuses on enhancing the reliability of machine learning predictions through the integration of blockchain. 
It suggests a method to train the algorithms on blockchain, particularly focusing on Naïve Bayes algorithm. The authors tackled the absence of floating point data support in the Ethereum Virtual Machine by using Taylor series expansion for probability estimation in integer arithmetic. 
% This method reduced the gas costs at the expense of lower accuracy. 
% The proposed solution was implemented in Solidity on the Ethereum blockchain and achieved prediction accuracy comparable to that of traditional machine learning libraries like scikit-learn in Python.
ML2SC \cite{10634431} was recently introduced as an open-source translation mechanism from PyTorch to Solidity. It facilitates deployment of the inference stage of multi-layer perceptron models as smart contracts. 
% The deployed smart contracts perform the inference stage of the model. The inference stage in machine learning models consists of floating-point calculations, whereas the Ethereum Virtual Machine (EVM) only supports integer calculations. 
In order to address the difficulty in handling floating point operations on EVM, ML2SC uses PRBMath \cite{prbmath}, which is a high-precision fixed-point math library. However, it incurs high gas cost even for a small ML model and works for binary classification only.

All of the above work require individual manual effort to generate smart contract code corresponding to any given ML model. This is inherently not scalable and is also prone to errors. In contrast, in LMST, we harness the power of publicly available large language models through elaborate prompt engineering techniques focused on gas cost optimization - a methodology that can work for any kind of neural network architecture.

\section{LLM based ML to Solidity Code Translation}
\label{sec:LMST}

We now present the main aspects of LMST - our proposed methodology. 
In LMST, the ML model training step is done off chain using a standard framework like PyTorch. The trained model weights and the inferencing part of the PyTorch code are translated into  Solidity using GPT-4 from OpenAI - one of the most popular LLMs. Unlike ML2SC as described above, LMST supports multi-class classification. It also effectively handles the challenges associated with ML code deployment on Ethereum using the following steps.   
\begin{itemize}
    \item \textbf{Transaction Splitting:} The data upload process is divided into multiple transactions in order to avoid exceeding the gas limit.
    \item \textbf{Increased Gas Limit on Local Test Network:} The gas limit in a local Ganache network can be increased from the standard limit (15 $\times$ 10\textsuperscript{6} on Ethereum mainnet) to 6.75 $\times$ 10\textsuperscript{10} for accommodating the process of uploading weights and testing data.
\end{itemize}

One of the main areas of focus in LMST is optimization of gas cost. The gas requirements for uploading weights and the test data are shown in Table \ref{table:gas_requirements}. As is observed, the gas values are significantly greater than the standard limit on the Ethereum mainnet.

\begin{table}[t]
\caption{Gas Requirements for Model Deployment on Local Test Network}
\label{table:gas_requirements}
\centering
\begin{tabular}{|c|c|}
\hline
\textbf{Phase} & \textbf{Gas Consumption} \\
\hline
Uploading weights and biases & 7.3 $\times$ 10\textsuperscript{7} \\
\hline
Uploading test data & 2.8 $\times$ 10\textsuperscript{9} \\
\hline
\end{tabular}
\end{table}

% Despite implementing the proposed solutions, the multi-class classification task on smart contracts remained challenging. This was due to the output layer restriction and lack of support for the softmax activation function.

% \subsection{Code Translation Using LLMs}
% We explored the use of Large Language Models (LLMs) for translating the PyTorch model code to Solidity smart contracts in order to overcome the limitations faced in ML2SC. We used GPT-4 for code translation. We aimed to translate the code for multi-class classification and minimize the need for manual intervention in the translation process. The complete process is described in the following subsections.

\subsection{Prompt Engineering}
\label{subsec:prompteng}
While LLMs are trained on a huge corpus, for deriving maximum benefit out of them, one has to choose an appropriate prompt that guides the LLM to generate the desired output. This step, known as prompt engineering, was meticulously followed in LMST. We started with a naïve prompt that aimed to guide the LLM in generating a Solidity code structurally resembling the code generated by the ML2SC translator but adapted for multi-class classification tasks. It was observed that the code generated by LLM was structurally similar to the code given by the ML2SC translator but had multiple syntax errors preventing compilation and requiring manual corrections to fix them.

These observations emphasized the need for improving the prompt in order to reduce syntax errors and enhancing translation efficiency. We, therefore, modified our prompt to include only the PyTorch code of the target model without any sample conversion. The objective was to assess the LLM's capability to translate the code and generate a Solidity smart contract without additional reference. Moreover, the prompt included instructions for dividing the phase of uploading weights and testing data into multiple smaller batches in order to avoid exceeding the gas limit.

\begin{figure}[h]
\centering
\includegraphics[height=0.2\textheight]{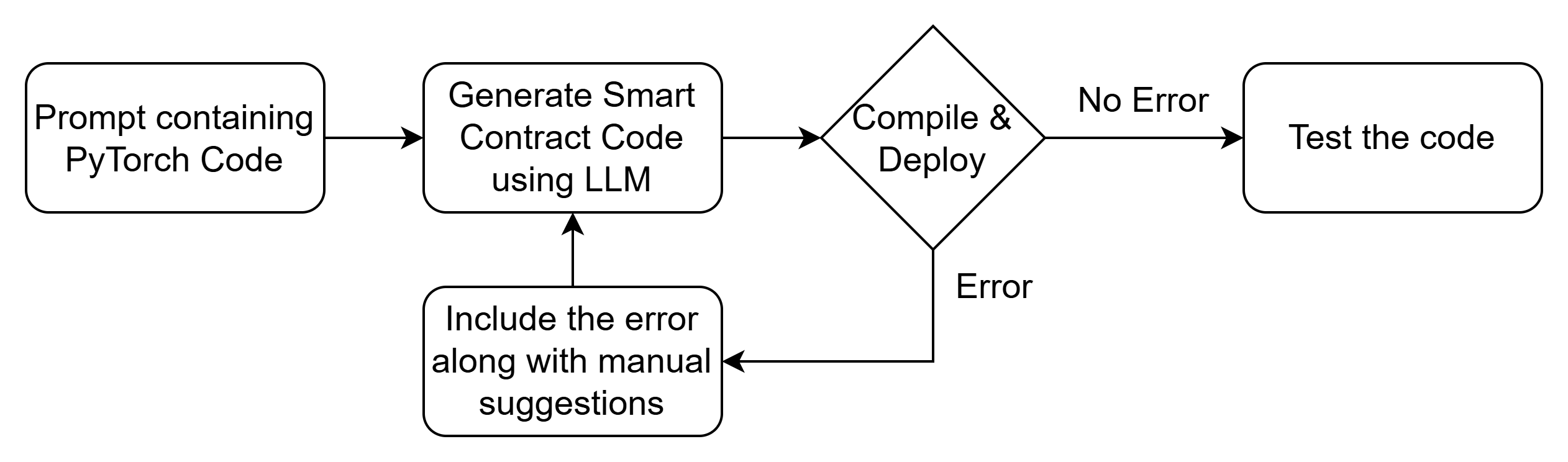}
\caption{LMST Overall Process Workflow}
\label{fig:Translation_Workflow}
\end{figure}

The overall process workflow of LMST in translating the model code using LLM is shown in Figure \ref{fig:Translation_Workflow}. We first write PyTorch code for a Multi-Layer Perceptron (MLP) model. This code is included in the prompt given to the LLM. The prompt also consists of several other instructions regarding the use of the PRBMath library for fixed-point computation and suggestions regarding transaction splitting in order to avoid exceeding the gas limit. When the LLM is supplied with this prompt, it generates a smart contract corresponding to the trained MLP model. Once the smart contract is generated, we try to compile it. If compilation fails, then the error along with a suggestion to resolve the error is given to the LLM, and the smart contract code is regenerated. Once the smart contract compiles successfully, we try to deploy the code on the local Ganache blockchain network. If deployment fails, the error is again supplied as a prompt along with suggestions to the LLM. The process is repeated until the code successfully gets deployed on the blockchain. After deployment, we perform gas cost analysis for the smart contract. 
% This is the overall workflow for the code translation process.

\subsection{Gas Cost Optimization}
\label{subsec:gasopt}

\begin{figure}[t]
\centering
\includegraphics[height=0.5\textheight]{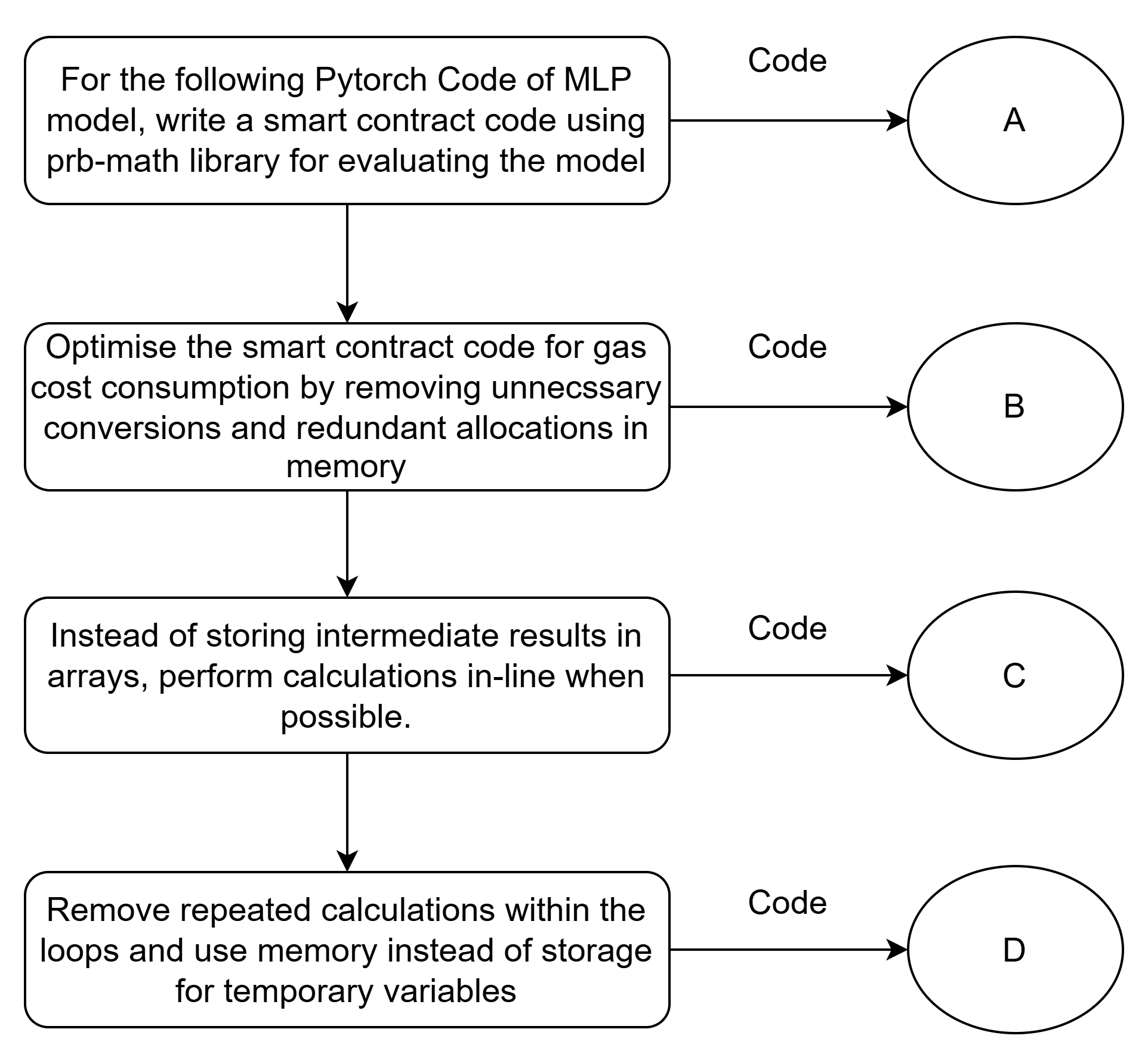}
\caption{Prompt Engineering Steps for Optimization Flow in LMST}
\label{fig:Optimisation_flow}
\end{figure}

In this sub-section, we explore different strategies for optimizing the gas used by LLM-generated smart contracts 
% Gas efficiency is an important factor to be considered while deploying and using smart contracts. It is crucial for computationally intensive tasks like machine learning model inference. The smart contract can be made more affordable and practical for real-world applications by reducing gas costs. 
through multiple steps. In each step, we create a prompt that focuses on a specific area of improvement and an optimization strategy. The optimization flow is shown in Figure \ref{fig:Optimisation_flow}. First, we supply the prompt obtained as output of Figure \ref{fig:Translation_Workflow} to the LLM. The translated code so obtained is labeled as A in Figure \ref{fig:Optimisation_flow}. The next prompt includes a suggestion to optimize the gas consumed by removing unnecessary conversions between PRBMath's fixed-point and Solidity built-in data types. It also suggests to avoid redundant memory allocations. The translated code corresponding to this prompt is labeled as B. The subsequent prompt asks the LLM to perform calculations in-line instead of creating additional variables for storing the intermediate values. The translated code corresponding to this prompt is labeled as C. In the final prompt, the LLM is asked to avoid repeated calculations within the loop and use memory data type for temporary variables. The translated code corresponding to this final prompt is labeled as D. The overall workflow for stepwise optimization of gas cost is now completed. 

\begin{figure}[t]
\centering
\fbox{\includegraphics[width=0.8\textwidth]{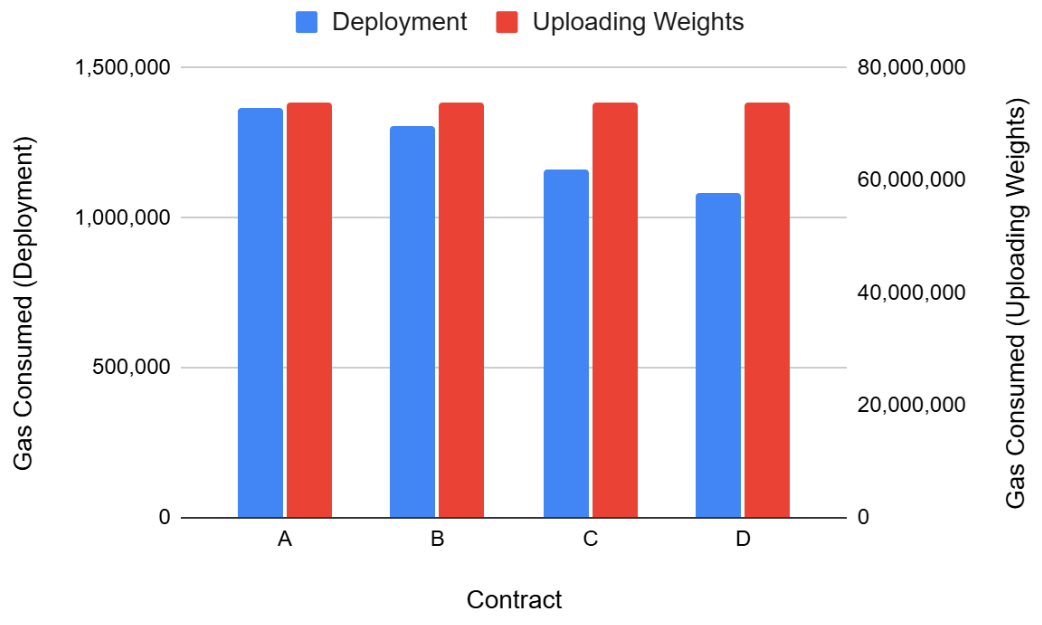}}
\caption{Optimization Step vs Gas Consumed}
\label{fig:Optimisation_vs_gas}
\end{figure}

We evaluated the smart contracts generated after every optimization step A-D by measuring the gas cost across two key stages: Deployment of smart contract, and Uploading of weights and biases. The image classification function is declared as \textit{view} and hence, does not modify the state of the blockchain. It, therefore, does not consume any gas. Figure \ref{fig:Optimisation_vs_gas} illustrates the gas consumed during deployment and uploading of weights after every optimization step. The deployment gas requirement decreases from optimization steps A to D, while gas consumed during uploading weights remains nearly constant over the stages. This is due to the fact that the gas consumed for weight uploading solely depends on the size of the weights. The overall gas cost progressively decreases from A to D. Each optimization step contributed to a meaningful reduction in the amount of gas consumed.

% \begin{figure}[]
% \centering
% \includegraphics[width=0.4\textwidth]{Pictures/deploy_op.png}
% \caption{Optimisation Step vs Gas Consumed during Deployment}
% \label{fig:Optimisation_vs_dgas}
% \end{figure}

% \begin{figure}[]
% \centering
% \includegraphics[width=0.4\textwidth]{Pictures/upload_op.png}
% \caption{Optimisation Step vs Gas Consumed during uploading weights}
% \label{fig:Optimisation_vs_ugas}
% \end{figure}

% \begin{figure}[]
% \centering
% \includegraphics[width=0.4\textwidth]{Pictures/total_op.png}
% \caption{Optimisation Step vs Total Gas Consumed}
% \label{fig:Optimisation_vs_tgas}
% \end{figure}

\section{Evaluation and Analysis}
\label{sec:eval}
In this section, we present the details of our experiments and the results obtained.
% \subsection{Dataset Description}
% \label{subsec:datasetdesc}
For multi-class classification, we chose the MNIST dataset \cite{lecun2010mnist}, which consists of handwritten digit images. Each image is labeled from 0 to 9, and the dimension of each image is 28$\times$28 pixels. A total of 60,000 training images are present in the dataset. 
% During our evaluation, we used a subset of 40 test images to keep the inference time low while assessing the model accuracy.

% \subsection{Model Description}
% \label{subsec:modeldesc}
We used a multi-layer perceptron for training and testing with the MNIST dataset. The MLP has 784 neurons in the input layer corresponding to the 28$\times$28 pixel input images and one hidden layer with 4 neurons. The output layer has ten neurons representing the 10 possible classes. This MLP model was trained on the complete data for 30 epochs using mini-batch gradient descent with a batch size of 64. The model achieved a test accuracy of 92.5\% in an off-chain setting.

% \subsection{Results}
% \label{subsec:results}
As mentioned in Section \ref{sec:LMST}, the output of LMST facilitates batch data uploading, thus avoiding the possibility of exceeding the gas limit. Accuracy of the model was computed over a subset of 1,000 test images. A comparison between on-chain and off-chain accuracy is shown in Table \ref{tab:accuracy_comparison}. It is seen that off-chain accuracy of the model is 86.0\%. On-chain accuracy, on the other hand, is 81.1\%. A key observation is that the three images incorrectly classified off-chain are also classified incorrectly on-chain. There is one extra mis-classification in on-chain evaluation, which was introduced due to fixed-point calculations. 
% Another important point to note is the time taken to classify a single image. 
% For off-chain, the time taken is less than 1 second, whereas for on-chain, the time taken is 5-10 seconds. 
% The total number of prompts required to get this working code was 8, and the total number of tokens utilized was 624.

\begin{table}[t]
\caption{Model Accuracy Comparison for 1000 Test Images}
\label{tab:accuracy_comparison}
\centering
\begin{tabular}{|c|c|}
\hline
\textbf{Scenario} & \textbf{Accuracy (\%)} \\
\hline
Off-chain PyTorch & 86.00 \\
\hline
On-chain Solidity & 81.10 \\
\hline
\end{tabular}
\end{table}
% The on-chain accuracy is less as compared to the off-chain accuracy. This is due to an error in the fixed-point calculation using the Prb-math library.
% \subsection{Model Complexity vs Gas Consumed}

We translated and deployed the smart contract for multiple MLP models with different numbers of neurons in the hidden layer. For each smart contract, the gas costs for deployment and uploading weights were measured. The structure of each model is denoted using the following notation: \textit{w}L\textit{x}N\textit{y}N..\textit{z}N where \textit{w} denotes the number of hidden layers (including the output layer), \textit{x},\textit{y}.. represent the number of neurons in the hidden layers and \textit{z} is the number of output layer neurons. Note that, for MNIST dataset, z=10 as it consists of ten digits. 

\begin{figure}[t]
\centering
\fbox{\includegraphics[width=0.8\textwidth]{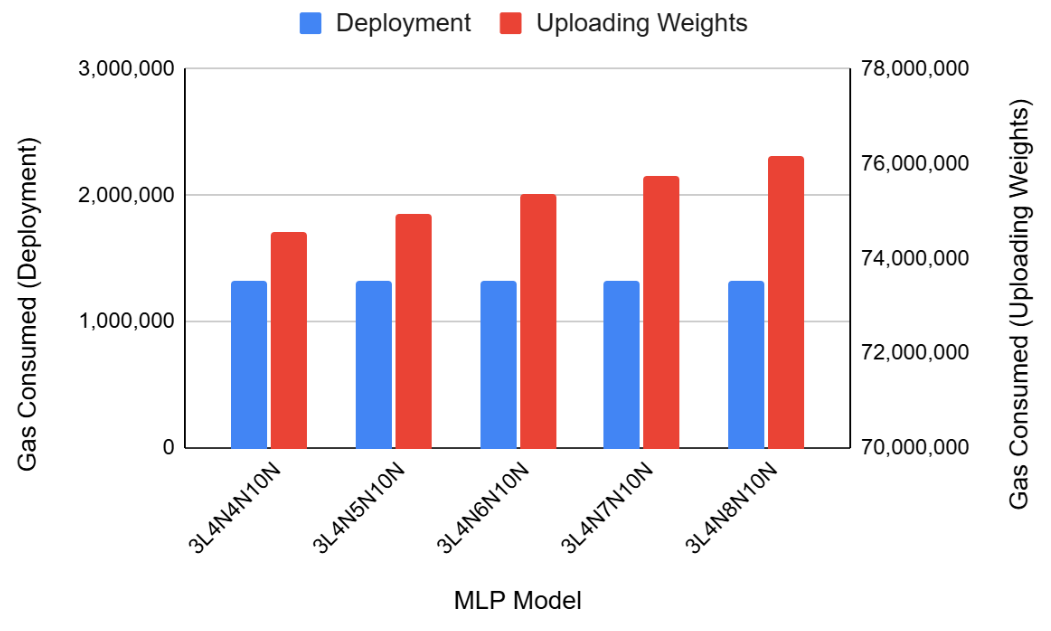}}
\caption{Model Complexity vs Gas Consumed}
\label{fig:Complexity_vs_gas}
\end{figure}

Figure \ref{fig:Complexity_vs_gas} illustrates both the variation in gas consumed during deployment of the contract and uploading of weights. It is observed that the gas requirement during deployment is constant for models with 3 layers while that during uploading of weights increases linearly with every neuron added. This is due to increase in the number of weights with each additional neuron. As we increase the number of neurons in the hidden layer, the overall gas consumed increases linearly.

% \begin{figure}[h]
% \centering
% \includegraphics[width=0.4\textwidth]{Pictures/deploy_mo.png}
% \caption{Model Complexity vs Gas Consumed during Deployment}
% \label{fig:Complexity_vs_dgas}
% \end{figure}

% \begin{figure}[h]
% \centering
% \includegraphics[width=0.4\textwidth]{Pictures/upload_mo.png}
% \caption{Model Complexity vs Gas Consumed during uploading weights}
% \label{fig:Complexity_vs_ugas}
% \end{figure}

% \begin{figure}[h]
% \centering
% \includegraphics[width=0.4\textwidth]{Pictures/total_mo.png}
% \caption{Model Complexity vs Total Gas Consumed}
% \label{fig:Complexity_vs_tgas}
% \end{figure}

% \subsection{Precision vs Model Accuracy:}
We next examine the effect of decreasing fixed-point precision on model accuracy. Table \ref{table:precision_correct_predictions} shows how the number of correct predictions (out of 40) varies by decreasing the fixed-point precision from 18 to 0. The number of correct predictions remains constant till 2 and then decreases. These results are due to the fact that while uploading the weights, we upload lower precision data, but the internal computations still use the 18 precision data type. This is because PRBMath does not support a data type with a precision of less than 18.

\begin{table}[t]
\caption{Precision vs Correct Predictions for 40 Test Images}
\label{table:precision_correct_predictions}
\centering
\begin{tabular}{|c|c|}
\hline
\textbf{Precision} & \textbf{Correct Predictions} \\
\hline
18 & 36 \\
10 & 36 \\
4  & 36 \\
2  & 36 \\
1  & 35 \\
0  & 6  \\
\hline
\end{tabular}
\end{table} 

% \subsection{Cost Analysis:}
Finally, we estimate the cost in both ETH and dollars for deploying the contract and uploading weights in the Ethereum mainnet. The values are shown in Table \ref{table:gas_cost}. The cost to deploy the contract is 0.054 ETH, equivalent to \$13.40. For uploading weights to the smart contract, the cost is 0.2949 ETH, which is equivalent to \$737.22. These estimates use the gas price of 4 Gwei and the price of 1 ETH to be \$2,500. The gas cost for classification is 0 as it is a \textit{view} function. 

\begin{table}[t]
\caption{Gas Usage and Cost Estimation}
\label{table:gas_cost}
\centering
\begin{tabular}{|l|c|c|c|}
\hline
\textbf{Phase}           & \textbf{Gas Used} & \textbf{Price in ETH} & \textbf{Price in \$} \\
\hline
Deployment               & 1,339,833                   & 0.0054                & 13.40                \\
Uploading Weights        & 73,721,648                  & 0.2949                & 737.22               \\
\hline
\end{tabular}
\end{table}

% \subsection{MNIST DApp}
We have developed a decentralized application where users can upload an image and get the predicted digit from the smart contract deployed on the Sepolia Ethereum test network. This dApp uses the smart contract developed by our LMST methodology as described in Section \ref{sec:LMST}. Any claims made towards accuracy of the ML model can thus be verified in a trustless manner.

\section{Conclusion and Future Work}
In this paper, we have proposed a methodology named LMST for deploying machine learning models on the Ethereum blockchain by utilizing code translation using GPT-4. Automating the translation process using LLMs is crucial for deploying any type of machine learning model. 
% A key aspect of the work is the optimization of the smart contracts by using targeted prompts to the LLM. Gas efficiency plays an important role in these computational heavy tasks. We successfully developed a DApp for facilitating multi-class classification on the MNIST dataset where users can upload an image and get the corresponding prediction. Also, the gas cost analysis done to deploy models provides significant insights into the financial implications of deploying machine learning models on the blockchain. 
We investigated how gas cost varies with respect to the model complexity. The approach showed similar accuracies for the model being executed on-chain and off-chain. 

We plan to extend our approach to more complex architectures such as Convolutional Neural Networks (CNN) and Long Short Term Memory (LSTMs). 
% Due to their computational intensity, efficient data flow and storage on-chain management will be required. Simultaneously, we need to maintain their performance characteristics. 
To improve the translation process from PyTorch to Solidity, we aim to implement RAG techniques, which combine retrieval-based methods with generative models. It enables more accurate and contextually relevant code translation. The translation process can also include external knowledge sources that provide examples for higher accuracy and further reduction in gas cost, especially for weight uploading. While Ethereum serves as a robust platform for deploying machine learning models, other Layer 1 and Layer 2 blockchain solutions can also be explored to understand their potential advantages.

% We also plan to integrate chain-of-thought prompting within the LLM framework to enhance the reasoning capabilities of the model during the code generation phase. This approach will encourage the model to articulate its thought process step by step. This will further lead to clearer and more logically structured outputs. Ethereum serves as a robust platform for deploying machine learning models. However, exploring other Layer 1 (L1) and Layer 2 (L2) blockchain solutions is essential to understand their potential advantages. L1 solutions such as Solana can reveal opportunities to improve scalability and reduce transaction costs. These platforms can also benefit in terms of speed and efficiency for machine learning deployments. We also aim to explore several Layer 2 solutions like Polygon, which can help decrease the gas cost issues in the Ethereum blockchain. We must assess these solutions to check how they can be integrated to enable faster transactions while maintaining security and decentralization.

% \clearpage
\bibliographystyle{IEEEtran}
\bibliography{Bibliography}

\end{document}